# Was the magnitude (M = 9.0R) of the mega-earthquake of Japan (11[th] of March, 2011) predictable? An analysis based on the Lithospheric Seismic Energy Flow Model (LSEFM).

## Thanassoulas[1], C., Klentos[2], V., Verveniotis, G.[3]

1. Retired from the Institute for Geology and Mineral Exploration (IGME), Geophysical Department, Athens, Greece.
   e-mail: thandin@otenet.gr - URL: www.earthquakeprediction.gr

2. Athens Water Supply & Sewerage Company (EYDAP),
   e-mail: klenvas@mycosmos.gr - URL: www.earthquakeprediction.gr

3. Retired Ass. Director, Physics Teacher at 2[nd] Senior High School of Pyrgos, Greece.
   e-mail: gver36@otenet.gr - URL: www.earthquakeprediction.gr

**Abstract**

The Tohoku EQ (11[th] of March, 2011, M = 9.0) in Japan falsified the proposed EQ magnitude range (M = 7.0 ÷ 8.5) of the same seismogenic regional area that had been determined by the compiled hazard maps, study of historical data, or other probabilistic methods while a larger magnitude (M > 9.0) had been proposed for all subduction zones. The observed discrepancy between the proposed EQ magnitude range and the actual one of the Tohoku EQ is studied in this work in terms of the cumulative seismic energy release of the study area and by the use of the Lithospheric Seismic Energy Flow Model (LSEFM). The results indicate that the Tohoku mega-earthquake magnitude could be predicted quite accurately provided that a long past seismic history had been available for use by the LSEFM procedure. Moreover, the presence, of the missing historic 1855 EQ (7.0< M <8.0) from seismic catalogs, was predicted backwards by the LSEFM method and its existence was verified by the Ishibashi (2004) work on Japanese historic seismicity. The recurrence time of the Tohoku EQ is estimated as being at least as 100 years. It is proposed frequent monitoring of the Japanese area seismic potential by compiling regularly in time the corresponding seismic potential maps.

**Key words:** Tohoku earthquake, earthquake magnitude, lithosphere, cumulative seismic energy, mega-earthquakes, seismic potential maps.

## 1. Introduction.

On March 11[th], 2011 a very large (M = 9.0R) and destructive earthquake occurred at Tohoku area, Japan. The extended damages and large number of deaths (nearly 20.000) raised the issue of the great difference between the expected and observed earthquake magnitude. Actually, a magnitude of 7.7 was proposed in the official hazard map for the Tohoku area (Geller, 2011; Stein et al., 2011, 2011a; see more references in: Kagan and Jackson, 2011) while Jackson et al. (2011) presented a summary for the various expected magnitudes for the Tohoku area as follows: Nishenko, 1991: 7.6 (characteristic), 8.1 (historical, 1611); Minoura et al., 2001: 8.3, 200x85 km, based on 869 Jogan Tsunami; Ruff and Kanamori, 1980: 8.2 (historical, age and plate-rate systematics); Bird and Kagan, 2004: Corner magnitude 9.6 for all subduction zones.; Koravos et al., 2006: 7.0< Mmax <8.0, based on historical earthquakes since 599 AD.; Annaka et al., 2007: 8.5 based on historical earthquakes since 1611.; Stein and Okal, 2011: 9.0+ for all subduction zones. In order to avoid lengthy references in existing methods and their problems for the estimation of the magnitude that is expected in a seismogenic area, we present the key - points from a Power Point presentation from Jackson et al. (2011).

Title: Estimating and testing earthquake magnitude limits

Quick Summary:
- We don't know what limits earthquake size.
- Maximum magnitude unknowable. "Maximum credible earthquake" totally subjective. Magnitude with defined excedance probability in specified time is a more useful concept.
- Common methods for estimating magnitude limits include (a) historic records, (b) fault or plate boundary segment size, and (c) moment balance in fixed area. All require subjective choices. Methods (a) and (b) have failed many times.
- Comprehensive assessments over large areas, e.g. all subduction zones, are safest methods. Bird and Kagan (2004) applied moment balance to a handful of global tectonic regions and have yet to be embarrassed.
- Methodology developed by the Collaboratory for Study of Earthquake predictability, (CSEP), can be used to test hypotheses of excedance probability in specified time.

Ways to estimate size limits:
- History of earthquakes within region
  – May use proxy such as tsunami run-up
- Fault-length or – area scaling
  – May use largest fault in a region to characterize the whole region.
- Copying Mmax of region with similar tectonics
- Tectonic and seismic moment balance, assuming a magnitude – frequency relation.

Problems for estimating size limits:
- Data are limited, especially temporally.
- Choice of region and its size is arbitrary.
- Definition of segments and rupture termination are imprecise.

- **Objectives conflict**
  - Small region has inadequate data, large region may miss important tectonics
  - Different communities have biases: academics, developers, insurers, environmentalists.

**Conclusions:**
1. We don't know what limits earthquake size. It does not seem to be segment boundaries, pre-existing fault extent, plate boundary triple junctions, or crustal thickness.
2. Efforts to estimate maximum magnitude based on geographically specific earthquake histories or tectonic situations tend to underestimate because of limited observations.
3. Some global or generic estimates (e.g. all subduction zones) haven't yet been falsified, but constructing facilities to withstand the implied upper limits could be impossibly expensive.
4. Unconditional maximum magnitude is not a scientifically meaningful concept, as it can't be tested in a finite time. Time limited probabilistic hypotheses are more useful and could possibly be testable.
5. Simple hypotheses that can be applied over much of the globe can be tested in a reasonable time; more complicated ones can't.
6. A CSEP experiment could fairly test some forecast probabilities of earthquakes over 7.5, 8.0, 8.5, 9.0 and 9.5 globally for 5 or 10 year period.

The problem of max magnitude estimation can be treated in an entirely different approach, in terms of energy conservation principle of physics. Since the magnitude of an earthquake is related to the released seismic energy during the seismic event, a different question could be posed as follows: what is the max amount of seismic energy to be released? The answer to this question is very simple: the max seismic energy to be released is what has already been stored in the seismogenic area since its last large seismic event. So, the new question is now: is it possible to determine that amount of stored seismic energy? The answer is "yes" by applying the Lithospheric Seismic Energy Flow Model (LSEFM). Its theoretical part with application on real earthquakes was presented by Thanassoulas et al. (2001), Thanassoulas (2007, 2008, 2008a), Thanassoulas et al. (2010, 2010a), and in URL: www.earthquakeprediction.gr. The latter method was applied on 18 EQs (1972 – 2001) of the Greek territory with Ms > 6.4R. The calculated magnitude deviated, in average, for only 0.067R from the seismologically calculated magnitudes. The LSEFM method was applied on the Park field EQ (2004) and on the Northridge EQ (1994) too. The Park field EQ deviated for only 0.06R (less) compared to the seismological magnitude, while the Northridge EQ deviated for only 0.29R (less) from the seismological one.

In this work the LSEFM method will be applied on the regional seismogenic Tohoku area in an attempt to test the predictability of the Tohoku EQ magnitude and to test the validity of the method in a different tectonic environment. It must be pointed out that the method does not rely on fault lengths, tectonics, crustal thickness etc. It is purely a generalized physical method that treats any seismogenic area as an "open physical system" that is capable of absorbing and releasing energy. The absorbed energy (mainly strain dynamic energy) charges, at a certain rate, the seismogenic area while at the same time the stored energy is released, at different variable rates, in small or large energy packets (EQs).

## 2. The large EQ (M = 9.0R) of Japan of the 11$^{th}$ of March, 2011

The location (after EMSC) of the Tohoku EQ is shown by a red star in the following map of fig. (1).

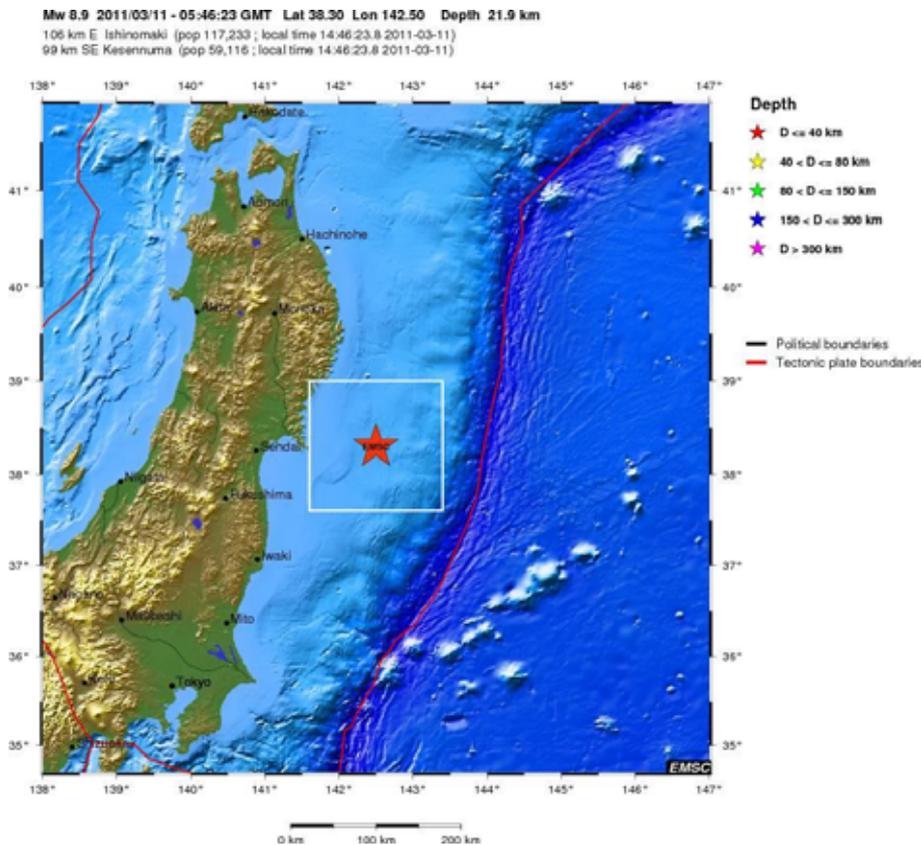

Fig. 1. Location (red star) of the large EQ (M = 9.0R) of Japan of the 11$^{th}$ of March, 2011 (after EMSC).

The adopted magnitude for the specific EQ varies according to the seismological observatory. In figure (1) the EMSC assigns a magnitude M = 8.9 while other seismological Institutions assign a magnitude of M = 9.1. In this work an average value of M = 9.0 has been adopted.

The basic requirements for the application of the LSEFM method are: a. definition of the areal extent of the open physical system and b. the existence of a long past seismicity history.

In the following figure (2), outlined by white lines, the adopted open physical system for the Tohoku EQ is shown. Its size is related to the estimated area assumed as affected during the preparation for an earthquake of such a magnitude.

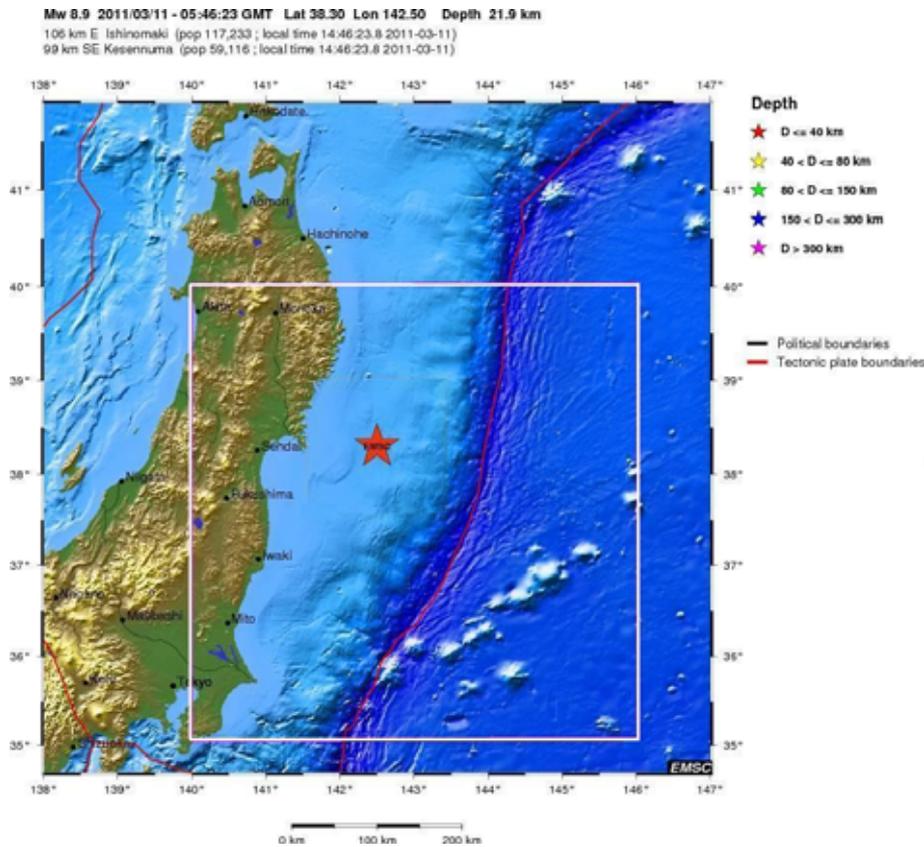

Fig. 2. Area extent of the adopted open physical system (white orthogonal)

The EQ data catalog which will be used is the one provided by the NOAA / NGDC. That catalog was processed and generated an EQ sub-catalog constrained with: Lat = 35 ÷ 40 and Lon = 140 ÷ 146. The resulted no. of EQs is 43 spanning from 1897 to 2011.

The results of the application of the LSEFM method derived by using the adopted open physical system and the corresponding seismicity of the same area (NOAA / NGDC sub-catalog) are presented in the following figure (3).

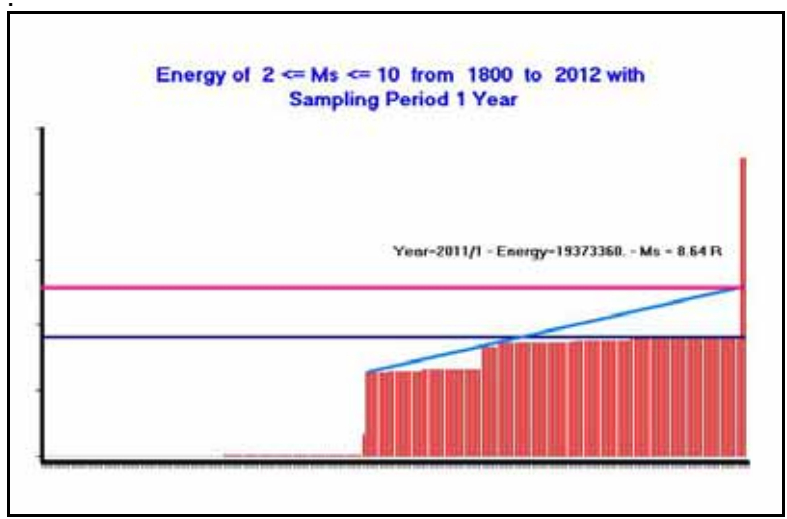

Fig. 3. Calculated magnitude (M = 8.64) for the Tohoku EQ by the LSEFM. Red vertical bars = cumulative seismic energy in year's samples. Black (lower) horizontal line = lower "lock" charge state of the seismogenic area, Inclined blue line = theoretical normal (cumulative) seismic energy release as time function, Red (upper) horizontal line = theoretical normal cumulative seismic energy release level that corresponds to the Tohoku EQ occurrence time. The far right cumulative energy release peak corresponds to the Tohoku EQ actual cumulative seismic energy release.

The calculated magnitude (M = 8.64) is: larger from all estimations made by statistical - seismological or historical analyses and: lower from what is expected for the subduction zones. It is evident that the methodology provides wrong results (M = 8.64 instead of M = 9.0) because some data input in the process is wrong. The calculated magnitude is directly related to the inclination of the theoretical cumulative seismic energy release function (blue line). Its inclination in turn depends on the length of the EQ catalog and the presence of large seismic events in the past. Thus, in n attempt to test if the methodology could provide correct results with extra correct data, firstly we adjust the theoretical cumulative seismic energy release function to fit the actual real conditions for a M = 9.0 **magnitude EQ. In other words we apply backwards application of the LSEFM for fitting an M = 9.0 EQ. This operation is shown in the following figure (4). The normal (theoretical) cumulative seismic energy release as a function of time is defined by two peaks. The first one (A) is the 1898 (M = 8.7) energy peak while the second (B) is the Tohoku EQ peak (right side of graph).**

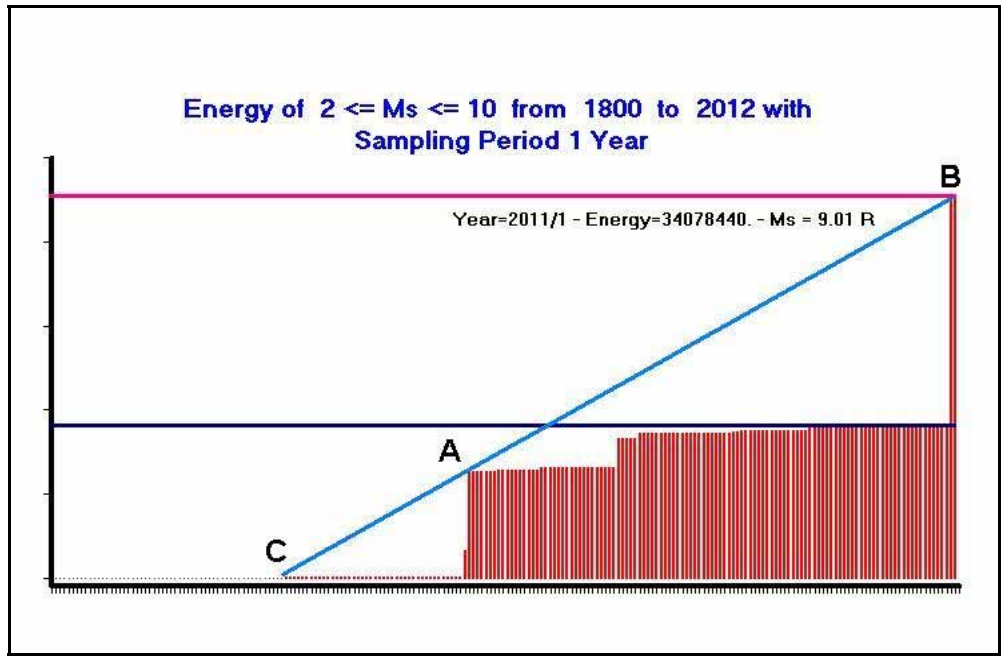

**Fig. 4. Backwards application of the LSEFM for fitting an M = 9.0 EQ. The normal (theoretical) cumulative seismic energy release as a function of time is defined by two peaks. The first one (A) is the 1898 (M = 8.7) energy peak while the second (B) is the Tohoku EQ peak (right side of graph). C represents the unknown past large seismic event.**

What is evident from figure (4) is that the LSEFM method fails to determine the correct magnitude just because the adopted EQ data file is shorter (backwards) than what is required in order to obtain a correct magnitude value since C past large seismic event is missing from the sub-catalog. In the next figure (5) a determination (occurrence time in years and magnitude) is made for the past (C) theoretical and yet unknown large EQ event which would fulfill the conditions required by the LSEFM for a obtaining a correct magnitude (M = 9.0) calculation for the Tohoku area.

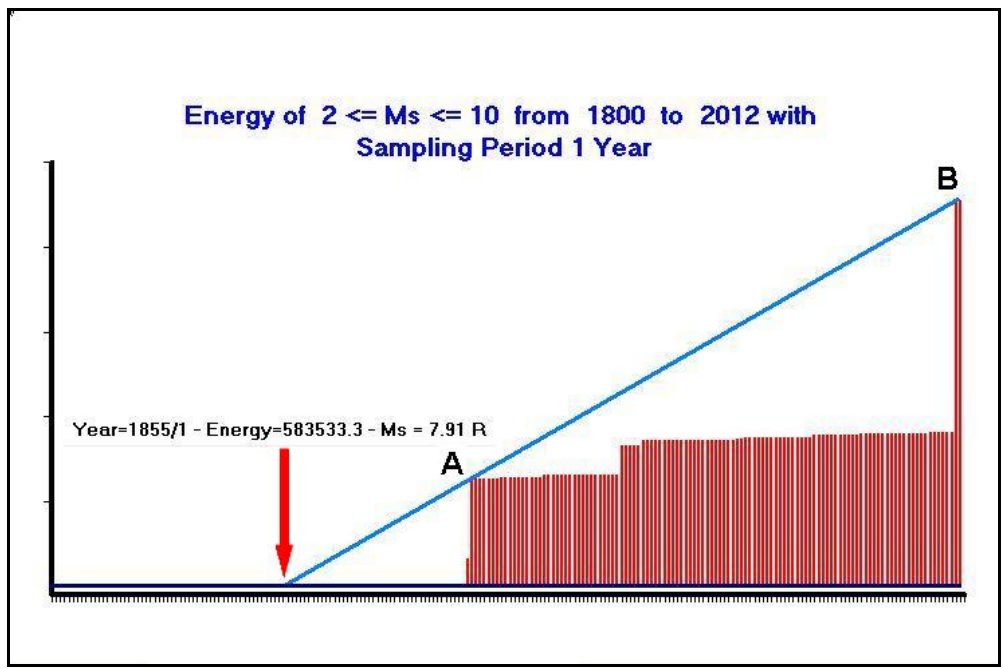

**Fig. 5. Past large EQ event (red arrow) determined for the Tohoku seimogenic area. It is calculated that an EQ of M = 7.91 on the year 1855 must exist in the past seismic history of the adopted open physical system so that figure (4) is justified. A = EQ on 1898 with M = 8.7, B = EQ on 2011 with M = 9.01.**

Since early EQs data are missing from the used NOAA / NGDC catalog, an investigation was performed in the seismological literature for the existence of papers related to historic earthquakes in Japan. The outcome of this investigation was a paper by Ishibashi (2004) that provided, in a map form, what we really needed for our work. This map is presented in the following figure (6).

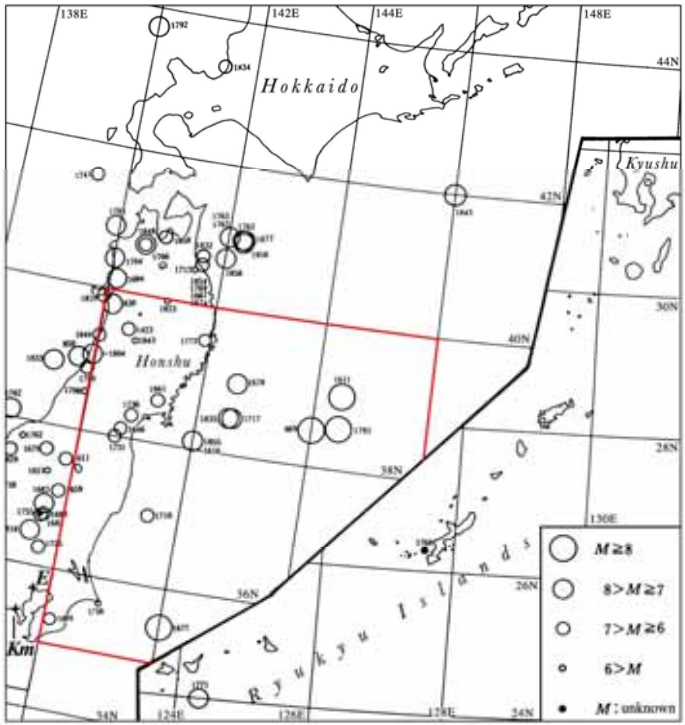

Fig. 6. Epicentral distribution of destructive earthquakes in Japan, since the beginning of history till 1872, as are shown by Usami (2003) for eastern and southwesternmost parts of Japan. The red lines indicate the extent of the adopted open physical system.

The next figure (7) presents an enlarged view of the area of interest. What is very important in this map is the fact that it verifies the presence of the expected (C) large EQ of the year 1855 with $7.0 \leq M \leq 8.0$ that occurred in the area extent of the adopted open physical system. That EQ is indicated by a red arrow.

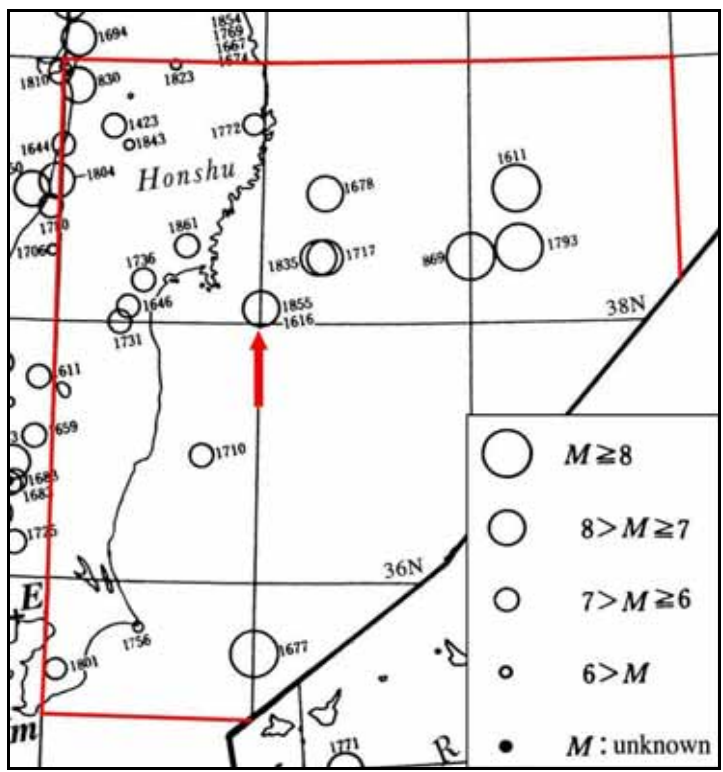

Fig. 7. Enlarged view of the area of interest (red orthogonal, Lat: $35^0 \div 40^0$, Lon: $140^0 \div 146^0$). Red arrow indicates the large EQ ($7.0 \leq M \leq 8.0$) event of 1855.

At a rather short distance (at NE direction) from the 1855 EQ location, another EQ of similar magnitude has occurred on 1835. That EQ deviates largely in time (20 years) from the determined year indicated in figure (5). Therefore it cannot be taken into account in applying the LSEFM.

For re-testing the applied LSEFM model with the new data, we modify the used EQ sub-catalog by removing the Tohoku EQ, since it is unknown yet, and by adding the 1855 EQ. The results of this operation are presented in the following figure (8).

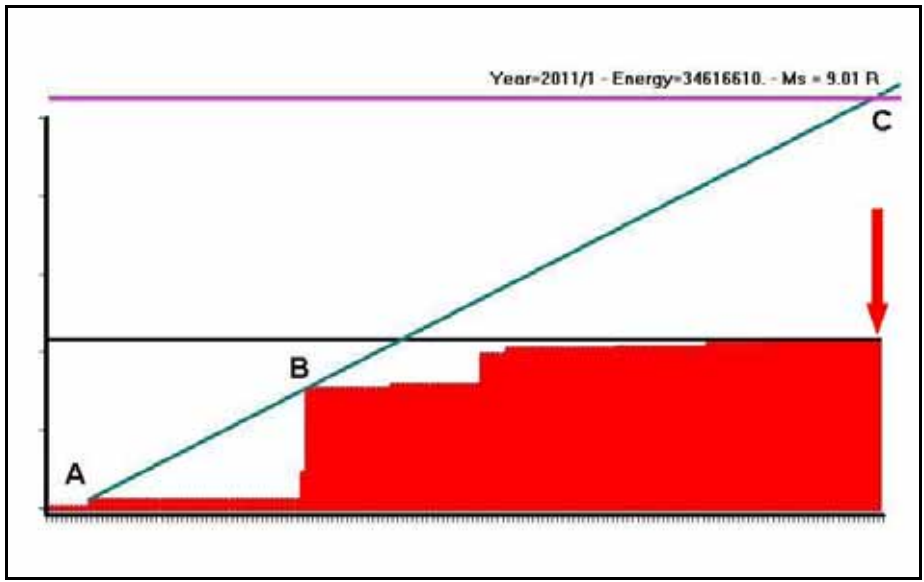

**Fig. 8. Calculation of the Tohoku EQ magnitude (red arrow) by using the modified EQ data file. A = EQ of 1855 with M = 7.91, B = EQ of 1898, with M = 8.7, C = EQ of 2011, with calculated M = 9.01 (Tohoku EQ).**

So far it has been demonstrated that if a most appropriate physical model and a longer (extending in the past) EQ catalog had been used, it could be possible to determine the max magnitude of the occurred large EQ in Tohoku area quite accurately. Furthermore, the LSEFM method can be used for the estimation of the time that will elapse until the next to come large EQ. The methodology is demonstrated in the following figure (9). What are known up to now are the theoretical cumulative seismic energy release time function (line A – B) and the time when recharging of the seismogenic area starts after the occurrence of the last large seismic event (B, Tohoku EQ). As a reference zero charge level for the cumulative seismic energy release is considered the horizontal black line that passes through B. Consequently, the max magnitude of a future EQ after B, at a specific time, is simply the observed cumulative energy difference between the normal cumulative seismic energy release (A - B line) and the zero reference level, determined at the time of hypothesized occurrence time of the future earthquake. From the graph of figure (9), after assigning an expected future magnitude of M = 9.01, the corresponding time of occurrence was determined as the year 2111.

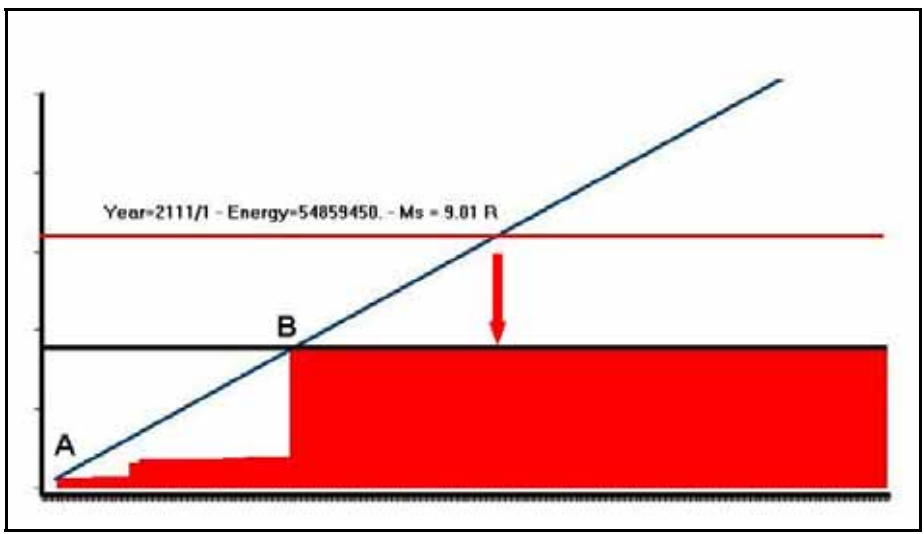

**Fig. 9. Determination of the occurrence time of the next, similar in magnitude to the Tohoku EQ (M = 9.01), seismic event. A = EQ of 1898 with M = 8.7, B = EQ of 2011 with M = 9.01, Red arrow = expected EQ of M = 9.01 that will occur theoretically on 2111.**

An assumption was made before the latter calculation. That is that no cumulative seismic energy release will take place in the seismogenic region before the occurrence of the expected in the future large EQ. In other words there will be no seismicity at all!! Obviously this is wrong. If a typical for the seismogenic area seismic activity is taken into account for a period of 100 years then the zero reference level of figure (9) will be raised above its present level and therefore the red arrow that indicates the time of occurrence of an EQ of M = 9.01 will be shifted further to the right in the time scale. Consequently, the 100 years interval, required for the recurrence of an M = 9.01 EQ, is a minimum theoretical recurrence time. Its actual expected recurrence cycle, at any time, can be monitored by studying (using the LSEFM) the seismic activity in time of the same seismogenic area. Recurrence rate, for large EQs of the same seismogenic area, have been presented from other researchers too. These recurrence rates span from 109 years (Kagan et al. 2011) to 260 - 880 years (Uchida and Matsuzawa, 2011) and to 500 - 1000 years (Simons et al. 2011).

An interesting application of the LSEFM method is its capability to calculate seismic potential maps. A seismic potential map is defined as the one compiled by the max magnitude values calculated spatially on a uniform grid over a large region. Such maps have been presented for the Greek territory by Thanassoulas, 2007; Thanassoulas et al. 2003; Thanassoulas et al. 2008; Thanassoulas et al. 2010. Seismic potential maps for the Greek territory have been compiled regularly at 5 years intervals from 1975 to 2010. It was revealed from these maps that the mapped seismic potential increased drastically within a 5 years period before any intense seismic activity took place and decreased very fast after it. Moreover, large EQs of the study period did occur in the areas which were characterized with large values of seismic potential that corresponded to EQ magnitudes of M > 6.0R.

We believe that it is worth for the Japanese Seismology Authorities to compile similar seismic potential maps for the entire regional Japanese seismogenic region. These maps will provide valuable updated information for the current seismic charge status of the Japan regional seismogenic area and preliminary information about the location of the most prone to large seismic events narrow zones.

## 3. Discussion - Conclusions.

The present work reveals some important outcomes:

a. It is possible to determine the seismic charge status of a seismogenic area without taking into account in a probabilistic sense its seismic history, tectonic conditions, known faults length or any other geotectonic or seismological parameters.
b. The seismogenic area in question can be treated as an "open physical system" that absorbs and releases energy in time no matter "how" and "why" it is utilized.
c. The seismogenic area is analyzed in terms of the Lithospheric Seismic Energy Flow Model (LSEFM) and its past seismic history. Consequently, a cumulative seismic energy release function of time is calculated through which the seismic charge status of the seismogenic area can be determined as a function of time and therefore, the max magnitude of any future EQ can be determined strictly by its expected occurrence time.
d. The Tohoku EQ max magnitude was determined correctly by applying the LSEFM method thus validating the methodology once more after the satisfactory results obtained by its application on a large number of large EQs from the Greek territory and the well-known Northridge and Park field EQs.
e. Moreover, the initially hypothesized 1855 large seismic event existence was backwards predicted and verified by papers related to the historic large and destructive seismic events of Japan.
f. The Tohoku EQ max magnitude recurrence time is determined as of 100 years at least.

Finally, the LSEFM methodology can be used to scan (with a pre-selected fixed grid size) a large area and to compile the corresponding seismic potential maps, in terms of max expected magnitude, at certain time increments. The latter has been applied for the Greek territory at 5 years intervals from 1975 to 2010 with very good results. Therefore, it is worth for the Japanese Seismology Authorities to compile similar seismic potential maps for the entire regional Japanese seismogenic region. These maps will provide valuable updated information for the current seismic charge status of the Japan regional seismogenic area and preliminary information about the location of the most prone to large seismic events narrow zones.